\documentclass[conference]{IEEEtran}
\IEEEoverridecommandlockouts
\usepackage{listings}
\usepackage{cite}
\usepackage{amsmath,amssymb,amsfonts}
\usepackage{algorithmic}
\usepackage{graphicx}
\usepackage{textcomp}
\usepackage{xcolor}
\usepackage{booktabs}
\usepackage{tikz}
\usepackage{subfigure}
\usepackage{booktabs}
\usepackage{todonotes}
\usepackage{color,xcolor,colortbl}
\usepackage{multirow}
\usepackage{enumitem}
\usepackage{url}
\usepackage[colorlinks,linkcolor=blue,citecolor=blue]{hyperref}

\def\BibTeX{{\rm B\kern-.05em{\sc i\kern-.025em b}\kern-.08em
    T\kern-.1667em\lower.7ex\hbox{E}\kern-.125emX}}

\begin{document}

\definecolor{codebackground}{rgb}{0.95,0.95,0.95}

\lstdefinestyle{solidity1}{
    basicstyle=\ttfamily\tiny,
    breaklines=true,
    frame=lines,
    numbers=left, 
    numberstyle=\tiny\color{gray},
    stepnumber=1, 
    numbersep=4pt, 
    backgroundcolor=\color{gray!10},
    commentstyle=\color{gray},
    keywordstyle=\color{blue},
    stringstyle=\color{red},
    keywords={pragma, contract, import, function, return, uint, address, public, view, returns, if, else, for},
    captionpos=b, 
    showstringspaces=false, 
    morecomment=[l]{//},
    morecomment=[s]{/*}{*/},
    morestring=[b]",
    morestring=[b]'
}

\lstdefinestyle{solidity2}{
    basicstyle=\ttfamily\small,
    breaklines=true,
    frame=lines,
    numbers=left, 
    numberstyle=\tiny\color{gray},
    stepnumber=1, 
    numbersep=4pt, 
    backgroundcolor=\color{gray!10},
    commentstyle=\color{gray},
    keywordstyle=\color{blue},
    stringstyle=\color{red},
    keywords={pragma, contract, import, function, return, uint, address, public, view, returns, if, else, for},
    captionpos=b, 
    showstringspaces=false, 
    morecomment=[l]{//},
    morecomment=[s]{/*}{*/},
    morestring=[b]",
    morestring=[b]'
}

\lstdefinestyle{json}{
    basicstyle=\ttfamily\small,
    showstringspaces=false,
    breaklines=true,
    frame=lines,
    backgroundcolor=\color{gray!10},
    stringstyle=\color{blue},
    keywordstyle=\color{purple},
    commentstyle=\color{gray},
    keywords={function_name, vulnerability, auditor, critic, correctness, severity, profitability},
    captionpos=b
}

\title{Large Language Model-Powered Smart Contract Vulnerability Detection: New Perspectives}

\makeatletter
\newcommand{\linebreakand}{%
  \end{@IEEEauthorhalign}
  \hfill\mbox{}\par
  \mbox{}\hfill\begin{@IEEEauthorhalign}
}
\makeatother

\author{
    \IEEEauthorblockN{Sihao Hu,  Tiansheng Huang, Fatih İlhan, Selim Furkan Tekin, Ling Liu \/}
    \IEEEauthorblockA{
        \textit{School of Computer Science}\\
        \textit{Georgia Institute of Technology}\\
        Atlanta, GA 30332, United States \\
       \{sihaohu, thuang, filhan, stekin6, ling.liu\}@gatech.edu
    }
}

\maketitle

\begin{abstract}

This paper provides a systematic analysis of the opportunities, challenges, and potential solutions of harnessing Large Language Models (LLMs) such as GPT-4 to dig out vulnerabilities within smart contracts based on our ongoing research.
For the task of smart contract vulnerability detection, achieving practical usability hinges on identifying as many true vulnerabilities as possible while minimizing the number of false positives. Nonetheless, our empirical study reveals contradictory yet interesting findings: generating more answers with higher randomness largely boosts the likelihood of producing a correct answer but inevitably leads to a higher number of false positives. To mitigate this tension, we propose an adversarial framework dubbed {\sc GPTLens} that breaks the conventional one-stage detection into two synergistic stages $-$ generation and discrimination, for progressive detection and refinement, wherein the LLM plays dual roles, \textit{i.e.}, {\sc auditor} and {\sc critic}, respectively. The goal of {\sc auditor} is to yield a broad spectrum of vulnerabilities with the hope of encompassing the correct answer, whereas the goal of {\sc critic} that evaluates the validity of identified vulnerabilities is to minimize the number of false positives. Experimental results and illustrative examples demonstrate that {\sc auditor} and {\sc critic} work together harmoniously to yield pronounced improvements over the conventional one-stage detection.
{\sc GPTLens} is intuitive, strategic, and entirely LLM-driven without relying on specialist expertise in smart contracts, showcasing its methodical generality and potential to detect a broad spectrum of vulnerabilities.
Our code is available at: \url{https://github.com/git-disl/GPTLens}.

\end{abstract}


\section{Introduction}

Smart contracts, commonly associated with cryptocurrency transactions on blockchains, were racked with financial losses up to billions of dollars due to vulnerability exploitation~\cite{zhou2023sok}. Due to the immutable nature once smart contracts are deployed, auditing acts an essential role in their development. Recently, generative Large Language Models~\cite{gpt3,openai2023gpt4,zhao2023survey} (LLMs) are rapidly emerging and reshaping the domain of software engineering~\cite{hou2023large}, facilitating tasks of code generation~\cite{codex}, code understanding~\cite{wang2023codet5+}, and code repair~\cite{paul2023automated}. Leveraging the capabilities of LLMs to empower smart contract auditing presents a promising application opportunity. In this paper, we envision the development of LLM-powered smart contract vulnerability detection techniques from a new perspective,  in tandem with a systematic analysis of the opportunities and challenges involved in this nascent research topic.

Compared to the existing representative analysis tools~\cite{feist2019slither,tsankov2018securify,kalra2018zeus,brent2018vandal,brent2020ethainter,mossberg2019manticore} developed in the past years, LLM-powered detection features some unparalleled advantages:

(1) \textit{Generality}: Existing tools like Slither~\cite{feist2019slither} require expert knowledge to design fixed-pattern detectors based on control-flow or data-flow, restricting them to specific types of vulnerabilities~\cite{dasp10}. In contrast, LLMs can emulate human linguistic understanding and reasoning and describe any type of vulnerability using natural language, allowing them to potentially detect a wider range of vulnerabilities, including those that are unknown or uncategorized a priori.

(2) \textit{Interpretability}: Generative LLMs can be utilized not only to detect vulnerabilities but also to offer intermediate reasoning about the detected vulnerabilities by following the chain-of-thought~\cite{chain-of-thought}. For programming and software engineering tasks, LLMs can provide insights into code understanding~\cite{wang2023codet5+} and even suggest code repair solutions~\cite{paul2023automated}. Such capabilities, if exploited intelligently, hold the potential to grant a heightened level of transparency and trustworthiness to the vulnerability detection process.

Nevertheless, certain limitations hinder LLMs from being exploited to their full potential:

(1) LLMs can produce a large number of false positives~\cite{david2023you}, resulting in a low precision and necessitate exhausting manual verification efforts. These false positive cases, categorized as factual errors or potential vulnerabilities, suggesting that they should be differentiated from the true vulnerabilities in terms of metrics beyond just correctness.

(2) LLMs, if used in a naive manner, tend to fail to uncover all vulnerabilities within the smart contract, leading to false negatives. These undetected vulnerabilities can be categorized into two primary groups: First, hard cases that exceed the ``cognitive ability'' of current LLMs; Second, vulnerabilities that are detectable but were missed because of the randomness of the generation. For the latter, our empirical study shows that instead of generating deterministic answers in one-shot, generating multiple answers with higher randomness (diversity) can largely boost the likelihood of the true answer being generated. Nevertheless, this strategy presents a Catch-22 dilemma~\cite{heller1999catch} as it inevitably introduces more false positives, \textit{i.e.}, the goal of detecting more true vulnerabilities is misaligned with the goal of reducing false positives.

To mitigate this tension, we propose {\sc GPTLens}, a framework that separates the conventional one-stage detection into two adversarial yet synergistic stages: \textit{generation} followed by \textit{discrimination}.

The primary goal of the generation stage is to enhance the likelihood of true vulnerabilities being identified (generated). With this goal in mind, we ask an LLM to play the role of multiple auditor agents, and each auditor generates answers (vulnerability and the reasoning) with high randomness, even though it could lead to a plethora of incorrect answers.
In contrast, the goal of the discrimination stage is to discriminate between true and false answers generated in the generation stage. To realize this, we prompt the LLM to play the role of a critic agent, which evaluates each identified vulnerability on a set of criteria, such as correctness, severity, and profitability, and assigns corresponding scores. Subsequently, {\sc GPTLens} ranks the answers by these scores to select the top-$k$ results.

The primary advantage of {\sc GPTLens} is that it resolves the Catch-22 dilemma present in one-stage detection, i.e., the conflicting goals between increasing the probability of generating the correct answer and reducing the number of false positives. Furthermore, {\sc GPTLens} is a \textit{pure} LLM-driven framework without resorting to any expert knowledge during the end-to-end vulnerability detection process.


We conduct preliminary experiments on 13 real-world smart contracts, all of which were reported to contain vulnerabilities in the Common Vulnerabilities and Exposures (CVEs) database~\cite{cve}. Experiments indicate that compared to the conventional one-stage detection which identifies true vulnerabilities in 38.5\% of contracts with the top-1 output, {\sc GPTLens} succeeds in 76.9\% of contracts. When comparing at the trial level, the accuracy for top-$1$ results rises from 33.3\% to 59.0\%. This enhancement is exciting as {\sc GPTLens} is simple and does not rely on any intricate design, suggesting its potential for broader application scenarios.

In summary, this paper makes three original contributions:
\begin{itemize}

    \item We provide a \textit{systematic} analysis of the advantages (opportunities) and challenges of LLM-powered smart contract vulnerability detection techniques.

    \item We introduce an \textit{innovative} framework, {\sc GPTLens}, constituted by two adversarial yet synergistic stages wherein the LLM takes on the roles of the auditor and critic agents respectively. 

    \item {\sc GPTLens} is \textit{simple}, \textit{effective} and \textit{purely} LLM-driven, eliminating the need for specialist expertise and showing the potential for generalization across a range of vulnerability types.
        
\end{itemize}


\section{LLM-powered Vulnerability Detection}

\subsection{Standard Detection Paradigms}

There are three standard prompting paradigms for vulnerability detection, \textit{i.e.}, binary prompting, multi-class prompting and open-ended prompting.

\begin{figure}[htbp]
\begin{tikzpicture}
  \node[draw, fill=gray!10, rectangle, rounded corners, inner sep=10pt] (box) {
    \begin{minipage}{0.45\textwidth}
      \textbf{Binary prompt:} You are a smart contract auditor. Review the following smart contract code in detail. Is the contract vulnerable to \{vul\_type\}? Reply with YES or NO only. \{contract code\}
    \end{minipage}
  };
  \label{binary_prompt}
\end{tikzpicture}
\vspace{-0.9cm}
\end{figure}

Existing works primarily follow the binary prompting paradigm~\cite{david2023you,gptscan}. In this paradigm, LLMs are prompted with the smart contract code and a specific vulnerability type~\cite{dasp10}, such as integer overflow/underflow, re-entrancy, or access control risk, and is expected to produce a binary YES or NO answer. These studies~\cite{david2023you,gptscan} also recommend including the definition or additional information about the vulnerability type to enhance performance. Given $n$ categories of vulnerabilities, the LLM service should be queried $n$ times for each smart contract.

\begin{figure}[htbp]
\vspace{-0.2cm}
\begin{tikzpicture}
  \node[draw, fill=gray!10, rectangle, rounded corners, inner sep=10pt] (box) {
    \begin{minipage}{0.45\textwidth}
      \textbf{Multi-class prompt:} You are a smart contract auditor. Here are \{$n$\} vulnerabilities: \{vul\_type$1$, vul\_type$2$, ..., vul\_type$n$\}. Review the following smart contract code in detail. Use 0 or 1 to indicate the presence of specific types of vulnerabilities, such as \{vul\_type$1$: 0, vul\_type$2$: 1, ..., vul\_type$n$: 0\}.      
    \{contract code\}
    \end{minipage}
  };
  \label{binary_prompt}
\end{tikzpicture}
\vspace{-0.6cm}
\end{figure}

An extension of binary prompting is multi-class prompting~\cite{chen2023chatgpt}, which requires LLMs to categorize identified vulnerabilities into multiple classes. Both binary and multi-class prompting fall under the category of \textit{close-ended} prompting, as they necessitate that the vulnerability categories be \textit{predefined}. Nevertheless, there always exist vulnerabilities that are either unknown or not categorized: Zhang et al.~\cite{zhang2023demystifying} found that 80\% of exploitable bugs remain undetected by existing analysis tools.

\begin{figure}[htbp]
\vspace{-0.2cm}
\begin{tikzpicture}
  \node[draw, fill=gray!10, rectangle, rounded corners, inner sep=10pt] (box) {
    \begin{minipage}{0.45\textwidth}
      \textbf{Open-ended prompt:} You are a smart contract auditor. Review the following smart contract code in detail and identify vulnerabilities within it.
    \{contract code\}
    \end{minipage}
  };
\end{tikzpicture}
\vspace{-0.5cm}
\end{figure}

In this paper, we propose a new prompting paradigm dubbed \textit{open-ended} prompting, which prompts LLMs to identify any potential vulnerabilities they think might be, and describe them in natural language without being constrained by predefined vulnerability names, theoretically enabling LLMs to recognize a broader range of vulnerability types.

\subsection{Advantages of LLM-powered Detection}

\textbf{Interpretability:} Beyond merely identifying vulnerabilities, we can ask LLMs to produce explanations for code, generate intermediate reasoning for vulnerability detection, generate examples of how to exploit identified vulnerabilities, and suggest code repair solutions. Such interpretability offers a new degree of transparency and trustworthiness, as we can gain insight into the step-by-step thought process~\cite{chain-of-thought} of LLMs. A case study in Listing~\ref{lst:19830} presents the explanations provided by GPT-4 for identified vulnerabilities.

\textbf{Generality:} Traditional smart contract auditing tools ~\cite{feist2019slither,brent2018vandal,kalra2018zeus,tsankov2018securify,mossberg2019manticore} have difficulty detecting unknown or uncategorized vulnerabilities since detectors are predesigned by human experts for fixed patterns of specific vulnerability types.

Existing AI-powered detection methods~\cite{qian2023cross,jeon2021smartcondetect,zhuang2021smart} also feature limited generality because they work in a supervised classification manner: vulnerabilities are classified into a fixed set of predefined categories based on known threats, which are used as the ground-truth to train a detection model.

For LLM-powered detection, although close-ended prompting also requires vulnerabilities to be predefined, open-ended prompting breaks this constraint. To retain the characteristic of generality, we adopt open-ended prompting throughout the paper. In our experiments, when prompting GPT-4~\cite{openai2023gpt4} with an open-ended prompt, it identifies ``condition logic error'' for CVE 2018-11411 and ``incorrect constructor name'' for CVE 2019-15079, which are semantically precise and fall outside of existing popular categorizations~\cite{dasp10}.


\textbf{Efficiency:} LLM services provide efficient online inference, making LLM-powered methods output results much faster than many traditional methods~\cite{chen2023chatgpt}. However, pre-training an LLM offline is prohibitively expensive in terms of both computational resources and time~\cite{touvron2023llama}.

\subsection{Limitations of Current LLM-powered Detection}
\label{sec:limitation}

Although LLM-powered detection offers promising advantages and despite the growing hype and claims regarding what LLMs can do, our empirical study has exposed some limitations inherent in current LLMs, which inhibit them from reaching their full potential in practice. Below, we discuss two primary limitations.

\begin{lstlisting}[style=solidity2,caption=Code snippet from the smart contract reported in CVE 2018-13836,label=lst:multiTransfer,framexleftmargin=-1pt,framexrightmargin=1pt]
function multiTransfer(address[] _addresses, uint[] _amounts) public returns (bool success) {
    require(_addresses.length <= 100 && _addresses.length == _amounts.length);
    uint totalAmount;
    for (uint a = 0; a < _amounts.length; a++)
        totalAmount += _amounts[a];
    require(totalAmount > 0 && balances[msg.sender] >= totalAmount);
    balances[msg.sender] -= totalAmount;
    for (uint b = 0; b < _addresses.length; b++) {
        if (_amounts[b] > 0) {
            balances[_addresses[b]] += _amounts[b];
            Transfer(msg.sender, _addresses[b], _amounts[b]);
        }
    }
    return true;
}
...
\end{lstlisting}

\subsubsection{Large number of false positives} 
The paramount challenge is that LLMs can produce a large number of false positives (FP), leading to exhausting manual verification efforts. A recent measurement study~\cite{david2023you} on project-level vulnerability detection demonstrates that GPT-4 can only identify 32 out of 73 vulnerabilities on 52 DeFi attacks, but produces 740 false positive cases, leading to an extremely low precision of 4.15\% ($Precision=\frac{TP}{TP+FP}$). A similar conclusion can be drawn from the results of Claude~\cite{anthropic2022claude}, which achieves a precision of 4.3\%. Another measurement study~\cite{chen2023chatgpt} shows that GPT-3.5 and GPT-4 achieve precisions of 19.7\% and 22.6\% respectively, in detecting the 9 most common categories of vulnerabilities.

In practice, we observe that false positives can primarily be broken down into two cases:

\begin{itemize}
    \item \textit{Factual error}: LLMs are insensitive to certain types of syntactic details, such as modifier statements, condition statements, error handling statements (\textit{require, assert, revert}), and event statements, especially when the number of input tokens is huge. For example, in Listing~\ref{lst:multiTransfer}, GPT-4 flags the \textit{multiTransfer} function for re-entrancy risk because it believes ``the function does not follow the Check-Effects-Interaction pattern, \textit{e.g.}, the state should not be updated before calling external contracts.'' Nonetheless, this false alarm stems from mistaking \textit{Transfer} as an external function call when it is actually an event statement for data logging without invoking external contracts.
\end{itemize}

\begin{lstlisting}[style=solidity2,caption=Code snippet from the smart contract reported in CVE 2018-10666,label=lst:10666,framexleftmargin=-1pt,framexrightmargin=1pt]
function setOwner(address _owner) returns (bool success) {
    owner = _owner;
    return true;
}
...
function uploadBalances(address[] addresses, uint256[] balances) onlyOwner {
    require(!balancesLocked);
    require(addresses.length == balances.length);
    uint256 sum;
    for (uint256 i = 0; i < uint256(addresses.length); i++) {
      sum = safeAdd(sum, safeSub(balances[i], balanceOf[addresses[i]]));
      balanceOf[addresses[i]] = balances[i];
    }
    balanceOf[owner] = safeSub(balanceOf[owner], sum);
  }
...
\end{lstlisting}

\begin{itemize}
    \item \textit{Potential vulnerability:} In another case, the identified risk does exist but remains unexploited, possibly because it is neither severe enough nor financially beneficial for attackers. In Listing~\ref{lst:10666}, GPT-4 highlights two vulnerabilities: the ``lack of access control'' in the \textit{setOwner} function, which allows anyone to call it, and the ``arbitrary balance manipulation'' in the \textit{uploadBalances} function, enabling the owner to set balances for any addresses arbitrarily, which could inflate the token supply. Despite both of them are correct vulnerabilities, only the former was exploited by the attacker and labeled as a CVE while the latter is considered as a false positive, since the formor is more severe and profitable. This observation indicates that vulnerability detection should consider not only correctness but also severity and profitability. Vulnerabilities detected in a smart contract should be ranked taking into account all these aspects.

\end{itemize}

While a recent effort~\cite{gptscan} seeks to mitigate the impact of false positives by utilizing sophisticated \textit{rules} for filtering, designing such rules demands expert knowledge and remains effective only for predefined vulnerability types. 


\subsubsection{Large number of false negatives} LLMs fail to detect a large portion of true vulnerabilities, resulting in a low recall ($Recall=\frac{TP}{TP+FN}$). As demonstrated in~\cite{david2023you}, GPT-4 and Claude-1.3 achieve recalls of 43.8\% and 35.6\% respectively on 52 DeFi attacks. In our experiments, we observe that false negatives can also be divided into two categories:

\begin{itemize}
    \item \textit{Hard cases} that are beyond the cognitive capabilities of current LLMs. It is reasonable to assume that certain vulnerabilities surpass the detection abilities of existing LLMs, including intricate logic issues that might elude even human auditors. To detect these hard cases, we expect more powerful LLMs in the future or more complicated designs, which will be discussed in Section~\ref{sec:future_direction}.

    \item Vulnerabilities that are detectable but \textit{undetected} due to the randomness of generation. As is known, GPT-like LLMs generate text by repeatedly estimating probability distributions for next positions across the vocabulary and sampling token-by-token~\cite{gpt3temp2021}. During generation, a hyper-parameter $t$ (temperature) controls the sharpness of the distribution~\cite{brown2020language}. Low randomness leads the LLM to generate more credible results, which outperforms than high randomness when the number of generated samples is small. However, even though high randomness leads to less credible results, it is more likely to generate a correct answer when the times of generation is huge. This observation cannot only be corroborated by our experiments (Section~\ref{sec:exp}), but also by the Codex paper~\cite{codex}: Figure~\ref{fig:codex_figure} shows the pass probability of the best result picked out of $k$ samples generated by Codex~\cite{codex} (a sibling of GPT-3) on a code generation task (HumanEval). 
    When the number of samples reaches 100, a higher temperature (0.8) outperforms a low temperature (0.2) with a huge margin (13 absolute percentage).

\end{itemize}

\begin{figure}[tbp]
\begin{center}
\includegraphics[width=7.5cm]{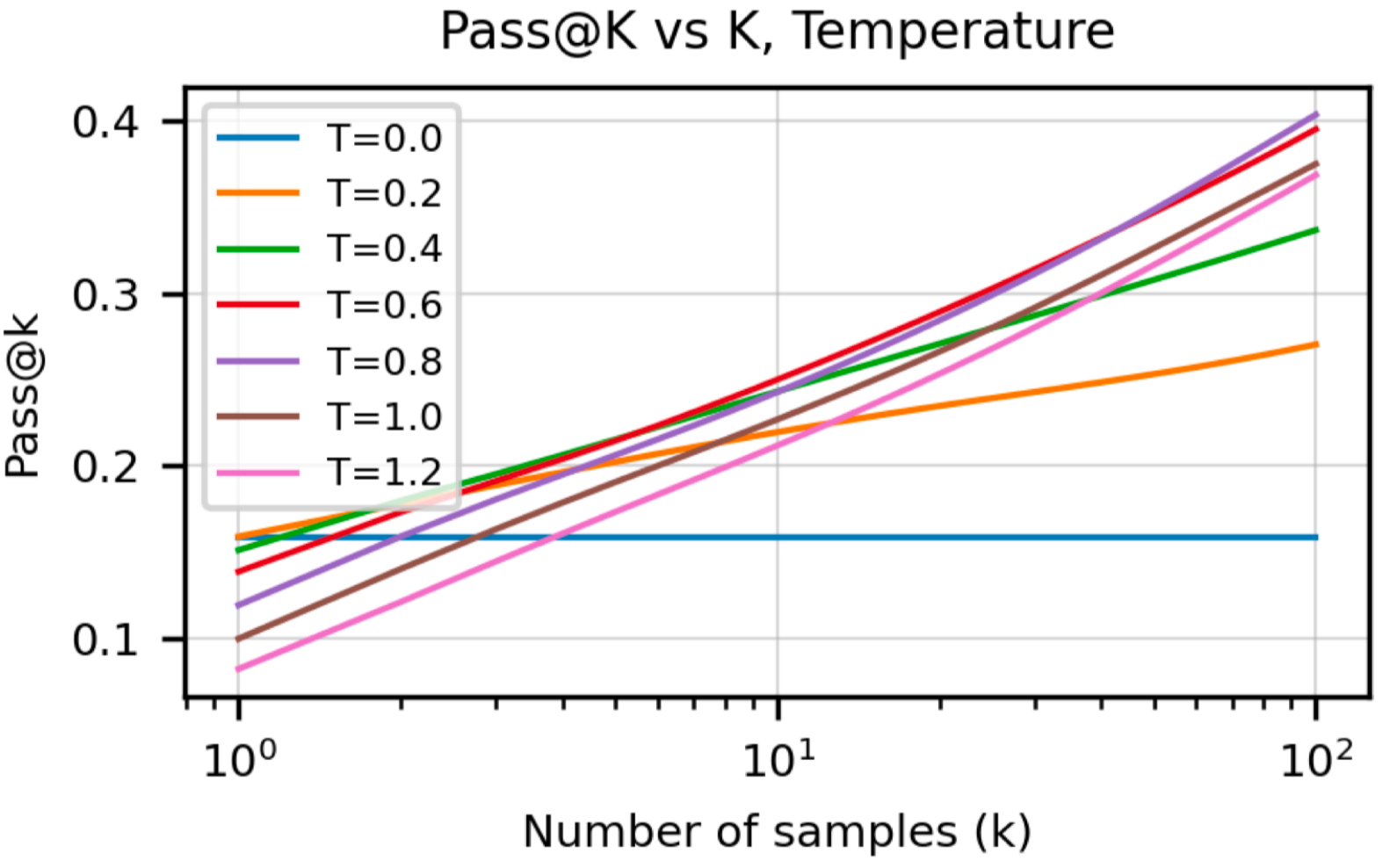}
\vspace{-5pt}
\caption{Pass@$k$ against the number of samples ($k$) \textit{w.r.t.} various temperatures reported in the Codex paper~\cite{codex}. Pass@$k$ is the probability of the best result out of $k$ generated samples, where the best sample is picked by an oracle (ground-truth knowledge). Higher the temperature $t$, higher the randomness of samples. When $t=0$ the LLM generates deterministic results. higher temperatures are better when the number of samples is large.}
\label{fig:codex_figure}
\end{center}
\vspace{-10pt}
\end{figure}  

Nevertheless, pass@$k$ in Figure~\ref{fig:codex_figure} is calculated utilizing an oracle (ground-truth knowledge), which is not available in real-world applications and generating more diverse answers inevitably leads to more false positives. Hence, how to identify more correct vulnerabilities without introducing more false positives is a challenge for LLM-powered detection.

\begin{figure*}[tbp]
\begin{center}
\includegraphics[width=18.5cm]{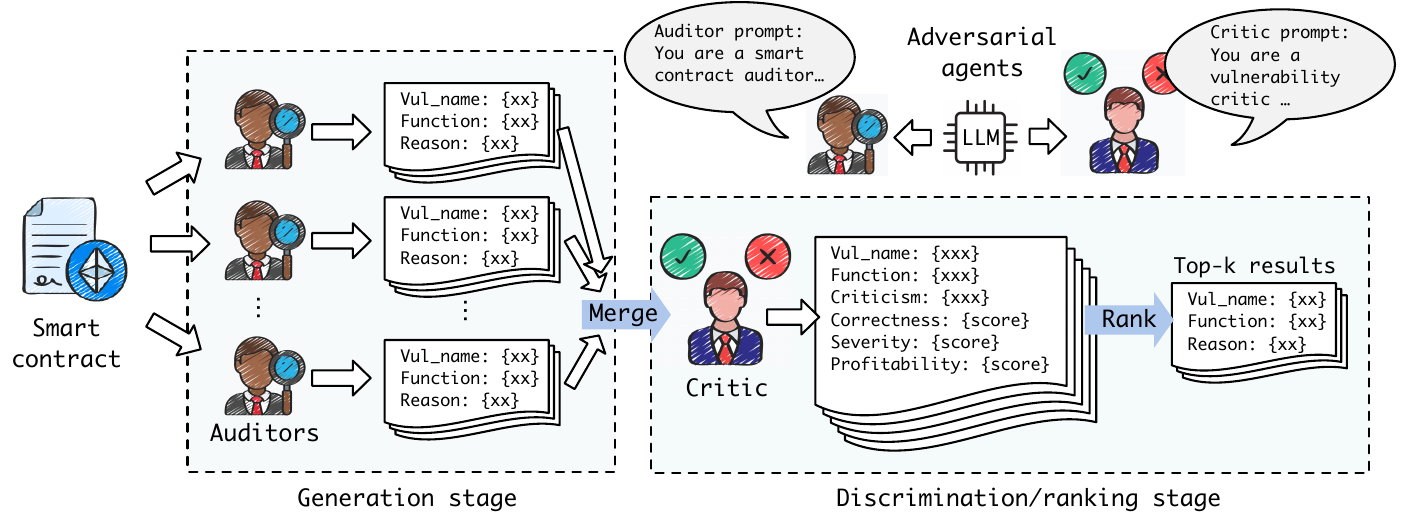}
\vspace{-20pt}
\caption{{\sc GPTLens}: an adversarial framework that breaks the conventional one-stage detection into the generation and discrimination stages.}
\vspace{-10pt}
\label{fig:framework}
\end{center}
\end{figure*}

\section{Two-Stage Adversarial Detection Framework}

\textbf{Goal:} Our goal is to present a simple, effective, and entirely LLM-driven methodology to shed light on the design of LLM-powered approaches. For every identified vulnerability, it should indicate the associated function and provide the reasoning.

\textbf{Motivations:} The design of {\sc GPTLens} is inspired by our previous analyses, summarized as follows:

\begin{itemize}
    \item[(1)] Open-ended prompting has good generalization across a wide range of vulnerabilities, including those unknown or uncategorized ones.

    \item[(2)] Reducing false positives is crucial for practical applications. False positives should be assessed not only for correctness but also for severity and profitability.
    
    \item[(3)] Generating a larger set of diverse samples can raise the likelihood of generating the correct answer, but it inevitably leads to more false positives.

\end{itemize}

The objective of identifying more correct vulnerabilities is \textit{in conflict} with the goal of reducing false positives in the current one-shot detection paradigm. To mitigate this tension, we break one-shot detection into two adversarial stages, \textit{i.e.}, the generation stage and discrimination/ranking stage. The idea of dividing one-stage into multiple stages is also employed in industrial recommendation systems~\cite{cao2022gift} to optimize respective goals at different stages.

Figure~\ref{fig:framework} shows the overall framework of {\sc GPTLens}, where an LLM plays the roles of two adversarial agents, \textit{i.e.}, the \textit{auditor} and the \textit{critic}, activated by different prompts in respective stages. 

In the generation stage, multiple auditors independently audit the smart contract code, generating identified vulnerabilities along with their associated functions and reasoning. The goal of this task is to yield a broad spectrum of answers, with the hope of encompassing the correct one.

In the discrimination stage, the identified vulnerabilities and their associated reasoning are scrutinized, evaluated and ranked by the critic agent, taking into account factors such as correctness, severity and profitability. The goal of this task is to simulate the role of an oracle, \textit{i.e.}, to precisely discern the correct answer and rank it above all other false positives. It is worth noting that the discrimination is not solely based on the identified vulnerability, but heavily leans on reasoning provided by auditors.

\textbf{Proof of feasibility:} Before diving into the details of {\sc GPTLens}, we raise a pertinent question: \textit{How is the critic agent able to discriminate between answers generated by the auditor agents if all the agents are played by the same LLM?}

The answer lies in the difficulty of two tasks: the generation task is more challenging than the discrimination task because the former demands the model to dig out all possible vulnerabilities within an entire smart contract project, while the latter only requires an assessment of recognized vulnerabilities at the function level. Moreover, the success of the adversarial framework has been previously corroborated by GANs~\cite{GAN}, wherein a very simple discriminator can be employed to train a sophisticated generator~\cite{radford2015unsupervised}.

To provide a clearer picture, in Figure~\ref{fig:prove_feasibility} we present the reasoning generated by an auditor along with the criticism from the critic for the code snippet in Listing~\ref{lst:10666}. The critic concludes that the vulnerability identified by the auditor is a factual error.

\begin{figure}[htbp]
    \centering
    \begin{tikzpicture}
    \node[draw, rectangle, rounded corners, inner sep=8pt] (box) {
    \begin{minipage}{0.45\textwidth}
      
      \textbf{Reasoning by the auditor}: There is a potential \textit{re-entrancy attack} within the \textit{MutiTransfer} function because it does not follow the Checks-Effects-Interactions pattern. The state is not updated before calling external contracts, which could potentially lead to re-entrancy attacks.     
    
      \vspace{0.2cm}
      \textbf{Criticism from the critic}: In the given \textit{multiTransfer} function, there doesn't appear to be any external calls before state changes. The function updates the balances appropriately before emitting a Transfer event. It's worth noting that events like \textit{Transfer} don't actually invoke external contract functions. They simply log data. Therefore, the vulnerability description ``Potential for Re-Entrancy Attack'' seems \textit{incorrect} based on the provided code.
      
\end{minipage}
};
\end{tikzpicture}
\vspace{-0.2cm}
\caption{Reasoning and criticism for code snippet in Listing~\ref{lst:10666}}
\label{fig:prove_feasibility}
\vspace{-0.3cm}
\end{figure}

\subsection{Generation Stage}

The goal of the generation stage is to identify a candidate set of vulnerabilities that includes the correct answer. Therefore, we set up $n$ auditors and initialize them with the auditor prompt presented in Figure~\ref{fig:auditor_prompt}. Each auditor is responsible for identifying up to $m$ vulnerabilities within the given smart contract code. The auditors' output strictly follows a JSON format for ease of parsing.
Temperature $t$ is set based on the hyper-parameters $n$ and $k$ since the optional value for $t$ varies depending on the number of samples. If $n$ is large, $t$ should also be set higher to prevent multiple auditors from generating similar answers.

\begin{figure}[tbp]
\begin{tikzpicture}
  \node[draw, fill=gray!10, rectangle, rounded corners, inner sep=10pt] (box) {
    \begin{minipage}{0.45\textwidth}
      \textbf{Auditor prompt:} You are a smart contract auditor, identify and explain severe vulnerabilities in the provided smart contract, ensuring they are real-world exploitable and beneficial to attackers. Include reasoning and corresponding function code as well. Output up to \{$m$\} most severe vulnerabilities. If no vulnerabilities are detected, output ``null''.
      
      \{contract code\} 

     Remember your output should adhere to the following format: \{json format\}.
    \end{minipage}
  };
\end{tikzpicture}
\vspace{-0.6cm}
\caption{Auditor prompt}
\vspace{-0.3cm}
\label{fig:auditor_prompt}
\end{figure}


\begin{figure}[htbp]
\begin{tikzpicture}
  \node[draw, fill=gray!10, rectangle, rounded corners, inner sep=10pt] (box) {
    \begin{minipage}{0.45\textwidth}
      \textbf{Critic prompt:} As a meticulous and harsh critic, your duty is to scrutinize the function and evaluate the identified vulnerabilities and reasonings with scores in terms of correctness, severity and profitability. Your criticism should include an explanation for your scoring. 
      
      \{output of auditors\} 

      Remember your output should adhere to the following format: \{json format\}.
      
    \end{minipage}
  };
\end{tikzpicture}
\vspace{-0.5cm}
\caption{Critic prompt}
\vspace{-0.5cm}
\label{fig:critic_prompt}
\end{figure}

\subsection{Discrimination/ranking Stage}

The role of the critic agent is to emulate an oracle, \textit{i.e.}, to discern the best answer from a multitude of false positives. To ascertain what is the best, we consider three distinct factors: correctness, severity and profitability, because although some false positives may be correct, they might possess a diminished level of severity or may not be profitable for attackers.

Concretely, the critic agent is activated using the critic prompt shown in Figure~\ref{fig:critic_prompt}, which directs the critic to evaluate the vulnerability, assign scores based on its reasoning, and provide explanation for these scores. Subsequently, we rank all vulnerabilities descendingly based on these scores and choose the top-$k$ vulnerabilities from the list as the output. Ideally, if the input list of vulnerabilities contains the ground-truth answer, the critic will place it at the forefront.

In our experiments, we employ only one critic agent to ensure that the criticism and scoring remain consistent across various vulnerabilities. Moreover, we set a low temperature value to reduce randomness.

\section{Experiment}
\label{sec:exp}

In this section, we validate the previous analyses and the efficacy of {\sc GPTLens} via experimental results.

\subsection{Experimental Settings}

\textbf{Dataset:} We collected the source code of 13 smart contracts from Etherscan\footnote{https://etherscan.io}, with each containing a reported vulnerability. We sourced the labels and descriptions for these vulnerabilities from the CVE database~\cite{cve}.

\textbf{Competitors:} The experiment involves six competitors under different settings. For methods, A, R, C, O represent Auditor, Random, Critic and Oracle. For parameters, $n$ denotes the number of auditors and $m$ denotes the maximum number of vulnerabilities identified by each auditor. The notation ($n$=2,$m$=3) means there are 2 auditors, and each auditor can generate up to 3 vulnerabilities. GPT-4 is adopted as the backend LLM. The descriptions for six competitors are as follows:
\begin{itemize}
    \item A($n$=1,$m$=1): One auditor identifies up to one vulnerability as the output (\textit{aka} one-stage detection).
    \item A+R($n$=1,$m$=3): One auditor identifies up to three vulnerabilities and randomly pick one as the output.
    \item A+C($n$=1,$m$=3): One auditor identifies up to three vulnerabilities, and the critic scores them. The vulnerability with the highest score is selected as the output.
    \item A+O($n$=1,$m$=3): One auditor identifies up to three vulnerabilities and an oracle is adopted to pick the best answer as the output.

    \item A+C($n$=2,$m$=3): Two auditors identify up to three vulnerabilities per each and the critic scores them. The vulnerability with highest score is selected as the output.
    
    \item A+O($n$=2,$m$=3): Two auditors identify up to three vulnerabilities per each and an oracle is adopted to pick the best answer as the output.
    
\end{itemize}

For all the auditors, the temperature $t$ is set to 0.7, while for the critic, $t$ is set to 0 to achieve confident and consistent scoring. It should be noted that the oracle leverages the ground-truth label information, therefore A+O only demonstrates an ideal performance. Due to the constraint on the token number per query, larger $n$ and $m$ are not tested in our experiments.

\begin{table}[tbp]
\centering
\caption{Hit times on 13 smart contracts logged in CVE database.}
\resizebox{0.49\textwidth}{!}{
\begin{tabular}{c|c|c|c|c|c|c}
\toprule
\textbf{Method} & A & A+R & A+C & A+O & A+C & A+O \\ 
\midrule
\textbf{Parameter} & $n1m1$ & $n1m3$ & $n1m3$ & $n1m3$ & $n2m3$ & $n2m3$\\
\midrule
\midrule
2018-10299 & 3 & 1 & 2 & 3 & 3 & 3 \\ 
2018-10666 & 3 & 1 & 3 & 3 & 3 & 3 \\ 
2018-11335 & 1 & 1 & 1 & 3 & 1 & 3 \\ 
2018-11411 & 0 & 2 & 2 & 3 & 2 & 3 \\ 
2018-12025 & 0 & 0 & 2 & 3 & 3 & 3 \\ 
2018-13836 & 2 & 0 & 1 & 1 & 0 & 2 \\ 
2018-15552 & 0 & 0 & 1 & 2 & 2 & 3 \\ 
2018-17882 & 0 & 0 & 2 & 2 & 3 & 3 \\ 
2018-19830 & 0 & 1 & 2 & 2 & 3 & 3 \\ 
2019-15078 & 0 & 0 & 0 & 0 & 0 & 0 \\ 
2019-15079 & 0 & 0 & 0 & 0 & 0 & 0 \\ 
2019-15080 & 3 & 1 & 2 & 3 & 3 & 3 \\ 
2018-18425 & 0 & 0 & 0 & 0 & 0 & 0 \\ 
\midrule
\midrule
Hit \# (CVE) &  5 & 6 & \textbf{10} & 10 & 9 & 10 \\
Hit ratio (CVE) &  38.5\% & 46.2\% & \textbf{76.9}\% & 76.9\% & 69.2\% & 76.9\% \\
\midrule
Hit \# (trail) & 13 & 7 & \textbf{18} & 25 & \textbf{23} & 29 \\ 
Hit ratio (trail) & 33.3\% & 18.0\% & \textbf{46.2}\% & 64.1\% & \textbf{59.0}\% & 74.4\% \\ 
\bottomrule 
\end{tabular}
}
\label{tab:comparison}
\end{table}

\subsection{Performance Comparison}

Each detection is conducted over three trials. For each trial, up to one vulnerability is selected as the output. A trial is deemed successful only if the function, vulnerability, and reasoning are all in alignment with the CVE report. The number of successful trials is presented in Table~\ref{tab:comparison}, where Hit \# (CVE) is the number of smart contracts for which the method correctly detected vulnerabilities at least once, and Hit \# (trial) represents the number of successful trials conducted across 13 smart contracts. 

From Table~\ref{tab:comparison} we can make several observations:
\begin{itemize}
    \item [(1)] At the trial level, A+R($n$=1,$m$=3) performs worse than A($n$=1,$m$=1), suggesting that generating more answers introduces more false positives. Nonetheless, at the contract level, A+R($n$=1,$m$=3) identifies some CVEs are not detected by A+R($n$=1,$m$=1) like 2018-11411 and 2018-19830, which implies that generating more answers increases the likelihood of generating correct answer. 
    
    \item [(2)] A more evident observation to support the above argument is that A+O($n$=1,$m$=3) works significantly better than A($n$=1,$m$=1).

    \item [(3)] A+C($n$=1,$m$=3) outperforms A+R($n$=1,$m$=3), indicating that the critic agent is very \textit{crucial} to discern true vulnerabilities from false positives introduced by generating more answers. A+C($n$=1,$m$=3) also works better than A($n$=1,$m$=1): the Hit ratio (CVE) increases from 38.5\% to 76.9\%, and the Hit ratio (trail) increases from 33.3\% to 46.2\%. 
    
    \item [(4)] Increasing the number of auditors can further improve the performance: compare A+C($n$=1,$m$=3) with A+C($n$=2,$m$=3), the Hit ratio (trail) increases from 46.2\% to 59.0\%. 
    
\end{itemize}


\subsection{Case Study} We provide a case study to demonstrate how {\sc GPTLens} performs by taking the smart contract code presented in Listing~\ref{lst:code19830} (see Appendix) as the input, with $n$ and $m$ set to 1 and 3 correspondingly. The outputs of both the auditor and critic are presented in Listing~\ref{lst:19830}.

In the generation stage, the auditor identifies three vulnerabilities and their associated reasoning: (1) ``Race condition'' in the \textit{approve} function (line 101-105); (2) ``Race condition'' in the \textit{transferFrom} function (line 93-100); (3) ``Unexpected behvaior'' in the \textit{UBSexToken} function (line 115-119). Although all explanations seem plausible, the first and second vulnerabilities are actually false positives. For the first vulnerability, the LLM overlooks the condition statement in line 102. For the second vulnerability, the LLM disregards the fact that Ethereum transactions are executed sequentially and atomically. As for the third vulnerability, its name differs from ``Constructor typo,'' which is labeled as the ground truth, Nevetheless, the reasoning and identified function are accurate.

In the discrimination stage, the critic uses the vulnerabilities, function code, and the auditor's reasoning as input and outputs criticism and scores with respect to correctness, severity, and profitability. As shown in Listing~\ref{lst:19830}, the critic determines that the reasoning for the first and second vulnerabilities is incorrect, assigning them low scores. Conversely, the critic recognizes that the reasoning and identification of the third vulnerability are accurate, assigning it the highest scores. Consequently, {\sc GPTLens} successfully identifies the genuine vulnerability within the smart contract.

As we can see, the entire detection process neither involves manually-defined heuristics and rules for fixed-pattern recognition nor requires a predefined set of vulnerability types, showing that {\sc GPTLens} holds the potential to generalize well across a wide range of vulnerabilities.


\begin{figure}[htbp]
\begin{lstlisting}[style=json,label=lst:19830,caption=A case study on CVE 2018-19830.]
{
    "function_name": "approve",
    "vulnerability": "Race Condition",
    "auditor": "This function does not reset the allowance before setting the new one. This can be exploited by the spender by front running the approver, allowing them to increase their allowance.",
    "critic": "The statement is correct in pointing out that this function does not inherently reset the allowance. However, the function does include a check to ensure that if the allowance is non-zero, the new value must be zero (and vice versa). Therefore, the risk is not as serve as stated.",
    "correctness": 3,
    "severity": 0,
    "profitability": 0
}
{
    "function_name": "transferFrom",
    "vulnerability": "Race Condition",
    "auditor": "The _allowance is fetched before the transfer is made, making it susceptible to a race condition. An attacker could call approve() in a transaction with a higher gas price, allowing them to increase the allowance before the transfer is executed.",
    "critic": "The statement is incorrect. The atomicity and sequential execution of Ethereum transactions ensure that there's no race condition between fetching _allowance and executing the token transfer.",
    "correctness": 2,
    "severity": 0,
    "profitability": 0
},
{
    "function_name": "UBSexToken",
    "vulnerability": "Unexpected Behaviour",
    "auditor": "The function name does not match the contract name. This indicates that this function is intended to be the constructor, but it is not. This means that anyone can call the function at any time and reset the totalSupply and owner of the token.",
    "critic": "The reasoning is correct. The function name does not match the contract name, indicating that it is intended to be the constructor but it is not. This means that anyone can call the function at any time and reset the totalSupply and owner of the token, which is a severe problem.",
    "correctness": 9,
    "severity": 9,
    "profitability": 9,
}
\end{lstlisting}
\end{figure}

\section{Future Directions}
\label{sec:future_direction}

To develop a general, practical LLM-powered smart contract vulnerability detection method, several directions can be explored in the future:

\textbf{Diversity in generation:} Increasing the diversity of the generation task helps to enhancing the finding of more true positives. It is benefical to explore innovative prompt engineering strategies and diversity metrics to encourage more generation and auditing diversity while maintaining the same number of generated samples.

\textbf{Consistency in discrimination:} Due to token count constraints, a large number of input vulnerabilities for the discrimination task need to be divided into multiple batches. This division can lead to scoring inconsistencies across different batches, even with a low temperature setting. In-context learning, using few-shot examples, could be explored to teach the LLM for more consistent scoring on novel observations and unseen events.

\textbf{Reasoning process optimization:} A popular direction to enhance the efficacy of LLMs is to design intricate reasoning processes that mimic the thought process of humans, such as chain-of-thoughts~\cite{chain-of-thought}, tree-of-thoughts~\cite{yao2023tree} and cumulative reasoning~\cite{zhang2023cumulative}, which can be adapted for the vulnerability detection task.


\textbf{Integrating Generative AI agents:} Generative AI agents~\cite{park2023generative,xi2023rise} employ LLMs as the core component and perform specific roles in various tasks like software development~\cite{qian2023communicative}. In this paper, we design only two synergistic roles for agents. Exploring additional roles for AI agents to achieve sophisticated functionalities, as well as designing how these agents collaboratively interact to solve complex detection tasks, are promising directions.


\textbf{Enabling external knowledge plug-in:} LLMs feature capabilities to use tools or call external APIs to expand their contextual knowledge~\cite{schick2023toolformer}. It would be intriguing to explore this functionality by allowing the LLM to autonomously determine when and what knowledge is beneficial for generating the correct answer for the vulnerability detection task.

\textbf{LLM-assisted tools:} Instead of serving as an end-to-end solution, LLM can be utilized as a tool to assist developers and auditors throughout the entire software engineering, including code generation~\cite{alphacode,codex}, code understanding~\cite{wang2023codet5+,shen2022benchmarking}, vulnerability detection~\cite{thapa2022transformer,wu2023effective} and code repair~\cite{paul2023automated,xia2023automated}, to name a few.


\section{Related work}

Various research efforts are dedicated to detecting vulnerabilities in smart contracts.

\textbf{Traditional methods:} Static analysis tools like Securify~\cite{tsankov2018securify}, Vandal~\cite{brent2018vandal}, Zeus~\cite{kalra2018zeus}, and Slither~\cite{feist2019slither} examine the source code without execution, aiming to detect potential vulnerabilities based on code patterns and structures. In comparison, dynamic analysis tools~\cite{jiang2018contractfuzzer,grieco2020echidna,wustholz2020harvey,zhang2019mpro} employ fuzz testing techniques that generates test inputs to identify anomalies during the actual execution of smart contracts, providing insights into runtime vulnerabilities. Symbolic execution tools like Manticore~\cite{mossberg2019manticore} and Mythril~\cite{mythril2023} investigate vulnerabilities across both bytecode and source code levels by examining all possible execution paths. Additionally, formal verification techniques, such as Verx~\cite{verx} and VeriSmart~\cite{verismart}, validate smart contracts against user-defined specifications, ensuring adherence to desired properties. 
 
\textbf{DL-powered methods:} Deep learning (DL)-based methods like sequence-based models~\cite{tann2018towards,qian2020towards}, CNN-based methods~\cite{sun2021attention}, graph neural networks-based methods~\cite{zhuang2021smart,liu2021combining,liu2021smart} are proposed to extract high-level representations to enhance the efficacy of vulnerability detection. Some hybrid methods~\cite{liao2022smartdagger,xue2022xfuzz,sendner2023smarter} combines deep-learning techniques with traditional methods. For examples, ESCORT~\cite{sendner2023smarter} and xFuzz~\cite{xue2022xfuzz} distill the outputs of traditional methods like Slither and Mythril to achieve good generality and inference efficiency. Some works~\cite{jeon2021smartcondetect,sun2023assbert} equipped with more advanced NLP techniques like BERT. A BERT-based approach~\cite{hu2023bert4eth} also demonstrates promising efficacy in Ethereum fraud detection tasks.

\textbf{Generative LLM-powered methods:} Very recent, some studies~\cite{david2023you,chen2023chatgpt} measure the performance of LLMs on the real-world datasets, suggesting that LLMs face precision-related challenges due to a high occurrence of false positives. GPTScan~\cite{gptscan} is introduced in an attempt to mitigate false positives by utilizing rule-based pre-processing and post-confirmation, which requires expert knowledge and extensive engineering efforts. In comparison, {\sc GPTLens} is more lightweight and entirely LLM-driven, making it general for a broader range of vulnerabilities.

\section{Conclusion}

This study provides a systematical analysis of harnessing generative LLMs for smart contract auditing, especially on the challenges of balancing the generation of correct answers against the backdrop of false positives. To address this Catch-22 dilemma, we present an innovative two-stage framework,  {\sc GPTLens}, by designing the LLM to play two adversarial agent roles: auditor and critic. The auditor focuses on uncovering diverse vulnerabilities complemented by intermediate reasoning while the critic assesses the validity of these vulnerabilities and the associated reasoning. Empirical results demonstrate that {\sc GPTLens} delivers pronounced improvements over the conventional one-stage detection and is entirely LLM-driven, which negates the dependency for specialist expertise in smart contracts and exhibits generalization to a broad spectrum of vulnerabilities.


\section{Acknowledgment}
This research is partially sponsored by the NSF CISE grants 2038029, 2302720, 2312758,  an IBM faculty award, and a grant from CISCO Edge AI program. 

\bibliographystyle{abbrv}
\bibliography{TPS23}

\appendices
\section{Source code of case study}
\begin{lstlisting}[style=solidity1,caption=Smart contract code reported in CVE 2018-19830,label=lst:code19830]
pragma solidity ^0.4.24;

library SafeMath {
    // ... (contents of the SafeMath library)
}

contract ERC20Basic {
    uint public totalSupply;
    function balanceOf(address who) constant returns (uint);
    function transfer(address to, uint value);
    event Transfer(address indexed from, address indexed to, uint value);
    function allowance(address owner, address spender) constant returns (uint);
    function transferFrom(address from, address to, uint value);
    function approve(address spender, uint value);
    event Approval(address indexed owner, address indexed spender, uint value);
}

contract BasicToken is ERC20Basic {
    using SafeMath for uint;
    address public owner;
    bool public transferable = true;
    mapping(address => uint) balances;
    mapping(address => bool) public frozenAccount;
    modifier onlyPayloadSize(uint size) {
        if (msg.data.length < size + 4) {
            throw;
        }
        _;
    }
    modifier unFrozenAccount {
        require(!frozenAccount[msg.sender]);
        _;
    }
    modifier onlyOwner {
        if (owner == msg.sender) {
            _;
        } else {
            InvalidCaller(msg.sender);
            throw;
        }
    }
    modifier onlyTransferable {
        if (transferable) {
            _;
        } else {
            LiquidityAlarm("The liquidity is switched off");
            throw;
        }
    }
    event FrozenFunds(address target, bool frozen);
    event InvalidCaller(address caller);
    event Burn(address caller, uint value);
    event OwnershipTransferred(address indexed from, address indexed to);
    event InvalidAccount(address indexed addr, bytes msg);
    event LiquidityAlarm(bytes msg);
    function transfer(address _to, uint _value) onlyPayloadSize(2 * 32) unFrozenAccount onlyTransferable {
        if (frozenAccount[_to]) {
            InvalidAccount(_to, "The receiver account is frozen");
        } else {
            balances[msg.sender] = balances[msg.sender].sub(_value);
            balances[_to] = balances[_to].add(_value);
            Transfer(msg.sender, _to, _value);
        } 
    }
    function balanceOf(address _owner) view returns (uint balance) {
        return balances[_owner];
    }
    function freezeAccount(address target, bool freeze) onlyOwner public {
        frozenAccount[target] = freeze;
        FrozenFunds(target, freeze);
    }
    function accountFrozenStatus(address target) view returns (bool frozen) {
        return frozenAccount[target];
    }
    function transferOwnership(address newOwner) onlyOwner public {
        if (newOwner != address(0)) {
            address oldOwner = owner;
            owner = newOwner;
            OwnershipTransferred(oldOwner, owner);
        }
    }
    function switchLiquidity(bool _transferable) onlyOwner returns (bool success) {
        transferable = _transferable;
        return true;
    }
    function liquidityStatus() view returns (bool _transferable) {
        return transferable;
    }
}

contract StandardToken is BasicToken {
    mapping(address => mapping(address => uint)) allowed;
    function transferFrom(address _from, address _to, uint _value) onlyPayloadSize(3 * 32) unFrozenAccount onlyTransferable {
        var _allowance = allowed[_from][msg.sender];
        require(!frozenAccount[_from] && !frozenAccount[_to]);
        balances[_to] = balances[_to].add(_value);
        balances[_from] = balances[_from].sub(_value);
        allowed[_from][msg.sender] = _allowance.sub(_value);
        Transfer(_from, _to, _value);
    }
    function approve(address _spender, uint _value) unFrozenAccount {
        if ((_value != 0) && (allowed[msg.sender][_spender] != 0)) throw;
        allowed[msg.sender][_spender] = _value;
        Approval(msg.sender, _spender, _value);
    }
    function allowance(address _owner, address _spender) view returns (uint remaining) {
        return allowed[_owner][_spender];
    }
}

contract BAFCToken is StandardToken {
    string public name = "Business Alliance Financial Circle";
    string public symbol = "BAFC";
    uint public decimals = 18;
    function UBSexToken() {
        owner = msg.sender;
        totalSupply = 1.9 * 10 ** 26;
        balances[owner] = totalSupply;
    }
    function () public payable {
        revert();
    }
}
\end{lstlisting}

\end{document}